\documentclass[prb,preprint,showpacs,eqsecnum]{revtex4}
\usepackage{graphicx}
\usepackage{bm}
\begin{document}
\title{Rapidly driven nanoparticles: Mean first-passage times and
relaxation of the magnetic moment}
\author{S.~I.~Denisov,$^{1,2}$ K.~Sakmann,$^{1}$ P.~Talkner,$^{1}$
and P.~H\"{a}nggi$^{1}$}

\affiliation{$^{1}$Institut f\"{u}r Physik, Universit\"{a}t Augsburg,
Universit\"{a}tsstra{\ss}e 1, D-86135 Augsburg, Germany\\
$^{2}$Sumy State University, 2 Rimsky-Korsakov Street, 40007 Sumy, Ukraine}


\begin{abstract}
We present an analytical method of calculating the mean first-passage times
(MFPTs) for the magnetic moment of a uniaxial nanoparticle which is driven by a
rapidly rotating, circularly polarized magnetic field and interacts with a heat
bath. The method is based on the solution of the equation for the MFPT derived
from the two-dimensional backward Fokker-Planck equation in the rotating frame.
We solve these equations in the high-frequency limit and perform precise,
numerical simulations which verify the analytical findings. The results are
used for the description of the rates of escape from the metastable domains
which in turn determine the magnetic relaxation dynamics. A main finding is
that the presence of a rotating field can cause a drastic decrease of the
relaxation time and a strong magnetization of the nanoparticle system. The
resulting stationary magnetization along the direction of the easy axis is
compared with the mean magnetization following from the stationary solution of
the Fokker-Planck equation.
\end{abstract}

\maketitle

\section{INTRODUCTION}

The problem of finding the statistical characteristics of the first, the
biggest, and the like, for sample paths of a stochastic process frequently
occurs in physics, biology, economics, and other sciences. \cite{HTB,Gar,Red} A
particularly prominent identifier is the mean first-passage time (MFPT), i.e.,
the average value of the random times taken by a random walker that starts out
from some initial state to reach another prescribed state in state space for
the first time. This quantity describes a large variety of noise induced
effects such as activation rates or reaction rates, the lifetime of metastable
states, the extinction of populations, or the extreme events in financial time
series, to name only a few. Unfortunately, the class of stochastic processes
for which the MFPT can be calculated explicitly is rather restricted. In fact,
the most general analytical results were obtained for one-dimensional discrete
or continuous  Markov processes that are \textit{homogeneous} in time.
\cite{PAV,Str,Goel,W2} However, these conditions of one-dimensionality and
time-homogeneity often represent an oversimplification. In particular, Markov
processes that describe \textit{time-depen\-dent} systems are usually not
homogeneous. Prominent examples are Brownian motors and ratchet-like stochastic
systems,\cite{Reim,L} as well as systems exhibiting Stochastic Resonance.
\cite{GHJM,H} Although a few advanced, approximate methods for the analysis of
periodically driven Markovian systems are available \cite{RTH,TH,T, LRH, MS,
SKTH, TL, SchTalHan} the development of new approaches for calculating the
MFPTs in such systems still presents an important challenge.

In this paper, we develop an analytical approach to the
\textit{two-dimensional} MFPT problem for a magnetic moment of a ferromagnetic
nanoparticle driven by a magnetic field which rapidly rotates in the plane
perpendicular to the easy axis of magnetization (up-down axis). The natural
precession of the magnetic moment always occurs in the counterclockwise
direction (when viewed from above). Therefore, its deterministic dynamics in
the up and down states differ.\cite{DLHT} For this  reason the stochastic
dynamics and thus the MFPTs in these states are different as well. In turn, the
difference in the MFPTs\- can drastically change the magnetic properties of
systems composed of nanoparticles. This was explicitly shown in the resonant
case, when the driving frequency coincides with the Larmor frequency of
precession.\cite{DLH} With this work we present a detailed analysis of this
phenomenon in the case of a rapidly rotating magnetic field, which was briefly
presented in Ref.~[\onlinecite{DSTH}], and apply the MFPTs for describing the
thermally activated magnetic relaxation in such systems.

The paper is structured as follows. In Sec.~II, we present the model and
introduce the stochastic Landau-Lifshitz equation together with the
corresponding forward and backward Fokker-Planck equations in the rotating
frame. In Sec.~III, we derive the general two-dimensional equations that define
the MFPTs for the driven magnetic moment of a uniaxial nanoparticle in the up
and down states. The analytical solution of these equations in the case of a
rapidly rotating magnetic field is carried out in Sec.~IV. In the same section,
to verify our method, we solve the effective stochastic Landau-Lifshitz
equations and calculate the MFPTs numerically. Some applications of the
obtained results are presented in Sec.~V. Here we study the features of
magnetic relaxation and steady-state magnetization induced by a rapidly
rotating field in nanoparticle systems. We summarize and discuss our findings
in Sec.~VI.

\section{MODEL AND BASIC EQUATIONS IN THE ROTATING FRAME}

Let us consider a simple model of a uniaxial ferromagnetic nanoparticle within
which the nanoparticle is characterized only by the anisotropy field $H_{a}$
and the magnetic moment $\mathbf{m}(t)$ of fixed length $|\mathbf{m}(t)| = m$.
This model is relevant for nanoparticles whose sizes do not exceed the exchange
length, i.e., the length scale below which the exchange interaction is
predominant. We assume also that perpendicular to the easy axis of
magnetization, which we choose as the z axis of a Cartesian coordinate system
$xyz$, a circularly polarized magnetic field $\mathbf{h}(t) = h(\cos\omega t,
\rho \sin\omega t,0)$ is applied. Here, $h = |\mathbf{h} (t)|$ is the field
amplitude, $\omega$ is the angular field frequency, and $\rho= -1$ or $+1$ for
clockwise and counter-clockwise rotation of $\mathbf{h}(t)$, respectively.

We take into account the influence of a heat bath by means of damping and the
presence of a thermal, Gaussian distributed magnetic field $\mathbf{n}(t)$,
possessing zero mean and the white noise correlations $\langle n_{\alpha} (t_1)
n_{\beta}(t_2) \rangle = 2\Delta \delta_ {\alpha \beta} \delta(t_2 - t_1)$.
Here, $n_{\alpha}(t)$ ($\alpha,\beta = x,y,z$) are the Cartesian components of
$\mathbf{n} (t)$, $\Delta$ is the intensity of the thermal field,
$\delta_{\alpha\beta}$ is the Kronecker symbol, $\delta(t)$ is the Dirac
$\delta$ function, and the angular brackets denote averaging with respect to
the sample paths of $\mathbf{n}(t)$. In this case the dynamics of
$\mathbf{m}(t)$ is Markovian and can be described by the stochastic
Landau-Lifshitz equation
\begin{equation}
    \frac{d}{dt}\mathbf{m} = -\gamma\mathbf{m}\times(\mathbf{H}_{\rm{eff}} +
    \mathbf{n}) - \frac{\lambda\gamma}{m}\,\mathbf{m}\times
    (\mathbf{m}\times\mathbf{H}_{\rm{eff}}),
    \label{st LL1}
\end{equation}
where $\gamma > 0$ denotes the gyromagnetic ratio, $\lambda(>0)$ is the
dimensionless damping parameter, $\mathbf{H}_{\rm{eff}} = -\partial W/\partial
\mathbf{m}$ is the effective magnetic field acting on $\mathbf{m}(t)$, $W$
denotes the magnetic energy of a nanoparticle, and the cross ($\times$)
indicates the vector product. The stochastic Landau-Lifshitz equation in the
form of (\ref{st LL1}) was introduced by Kubo and Hashitsume. \cite{KH} Another
form of this equation (with Gilbert's relaxation term \cite{Gil}) was employed
by Brown in his well-known paper.\cite{B} We note, however, that although the
solutions of these equations for a given realization of $\mathbf{n}(t)$ are
generally different, their statistical characteristics are the same \cite{GP}
(to within a renormalization factor for $\Delta$). At present, both forms of
the stochastic Landau-Lifshitz equation are used equally often. Although in the
Langevin equation (\ref{st LL1}) the noise enters in a multiplicative manner,
the resulting process actually is independent of the stochastic calculus if
restricted to the sphere $\mathbf{m}^2(t) =m^2$. This is so, because the
corresponding, noise induced drift terms are perpendicular to the sphere for
{\it any} interpretation of the stochastic differential equation (\ref{st
LL1}).\cite{hanggiHPA,HT,Ber,BerG} For the sake of definiteness, upon
performing nonlinear transformations of this stochastic differential equation
we shall consistently employ the Stratonovich interpretation.\cite{S} We next
perform such a nonlinear transformation by considering the dynamics of
$\mathbf{m}(t)$ in terms of spherical coordinates in a rotating frame.

Specifically, using the respective polar and azimuthal angles $\theta$ and
$\varphi$ of the magnetic moment, $\mathbf {m}(t) = m( \sin \theta \cos
\varphi, \sin \theta \sin \varphi, \cos \theta)$, the magnetic energy $W$
emerges as
\begin{equation}
    W = \frac{1}{2}\,mH_{a}\sin^{2}\theta - mh \sin\theta \cos(\varphi
-\rho \omega t).
    \label{W}
\end{equation}
The energy $W$  depends on $\varphi$ and $t$ only through the single variable
$\psi= \varphi - \rho \omega t$. Therefore, it is advantageous to introduce a
\textit{rotating} Cartesian coordinate system $x'y'z'$ (see also a similar
description of a noisy, periodically driven Van der Pol oscillator in
Ref.~[\onlinecite{AmJP}]), in which $\mathbf{h}(t) = h(1,0,0)$ and the
azimuthal angle $\varphi$ is changed by $\psi$. According to Eq.~(\ref{st
LL1}), in this coordinate system the equations for $\theta$ and $\psi$ can be
written in the dimensionless form as follows:
\begin{eqnarray}
    &\dot{\theta} = u(\theta,\psi) + \xi_{\theta}(\theta,\psi,\tau),&
    \nonumber\\[6pt]
    &\dot{\psi} = v(\theta,\psi) - \rho\Omega +
\xi_{\psi}(\theta,\psi,\tau). &
    \label{st LL2}
\end{eqnarray}
Here, an overdot denotes the derivative with respect to the dimensionless time
$\tau = \omega_{r}t$ with the Larmor frequency $\omega_{r} = \gamma H_{a}$, and
$\Omega = \omega/\omega_{r}$ is the corresponding dimensionless frequency of
the driving field. The functions $u(\theta,\psi)$ and $v(\theta,\psi)$ result
as
\begin{eqnarray}
    &u(\theta,\psi) = -\displaystyle\frac{1}{\sin\theta}
    \left(\lambda\sin\theta \frac{\partial}{\partial\theta} +
    \frac{\partial}{\partial\psi} \right)\tilde{W},&\nonumber \\[6pt]
    &v(\theta,\psi) = \displaystyle\frac{1}{\sin^{2}\theta}
    \left(\sin\theta\frac{\partial}{\partial\theta} - \lambda
    \frac{\partial}{\partial\psi}\right)\tilde{W}&
    \label{uv1}
\end{eqnarray}
with $\tilde{W} = W(\theta,\psi)/mH_{a}$ denoting the dimensionless magnetic
energy. The stochastic forces $\xi_{\theta,\psi}(\theta, \psi,\tau)$ are
determined as
\begin{eqnarray}
    &\xi_{\theta} = - \tilde{n}_{x'}(\tau) \sin\psi + \tilde{n}_{y'}(\tau)
    \cos\psi,&\nonumber \\[6pt]
    & \xi_{\psi} = \tilde{n}_{z}(\tau) - [\tilde{n}_{x'}(\tau) \cos\psi
    + \tilde{n}_{y'}(\tau) \sin\psi]\cot\theta, \quad&
    \label{xi1,2}
\end{eqnarray}
where $\tilde{n}_{x'}(\tau) = \tilde{n}_{x}(\tau) \cos\Omega\tau + \rho \tilde
{n}_{y}(\tau) \sin\Omega\tau$, $\tilde{n}_{y'}(\tau) = \tilde {n}_{y} (\tau)
\cos\Omega\tau - \rho\tilde{n}_{x}(\tau) \sin\Omega\tau$, and $\tilde{n}
_{\alpha} (\tau) = n_{\alpha}(\tau/\omega_{r})/H_{a}$ are the components of the
reduced thermal magnetic field. Using the statistical characteristics of
$n_{\alpha} (t)$, for these components we readily obtain $\langle \tilde{n}
_{\alpha} (\tau) \rangle = 0$ and $\langle \tilde{n}_{\alpha} (\tau_1)
\tilde{n}_{\beta} (\tau_2) \rangle = 2 \tilde{\Delta} \delta_ {\alpha \beta}
\delta(\tau_2 - \tau_1)$. Here, $\tilde{\Delta} = \Delta\gamma/H_{a}$ is the
dimensionless intensity of the reduced thermal field. With the help of the
Sutherland-Einstein relation,\cite{chaosHM} i.e., $\Delta = \lambda
k_{B}T/\gamma m$ wherein $k_{B}$ denotes the Boltzmann constant and $T$ is the
absolute temperature, it can be written also in the form $\tilde {\Delta} =
\lambda/2a$, where
\begin{equation}
    a = mH_{a}/2k_{B}T
    \label{a}
\end{equation}
is the anisotropy barrier height in the units of the thermal energy $k_{B}T$.
We note that in the purely deterministic case, when $\tilde {\Delta} = 0$, some
important features of the solution of Eqs.~(\ref{st LL2}) were studied in the
context of the nonlinear dynamics of $\mathbf{m}(t)$ and its
stability.\cite{BSM, DLHT}

Next we introduce the conditional probability density $P = P(\theta,\psi, \tau
|\theta',\psi',\tau')$ ($\tau \geq \tau'$) which presents the most important
statistical characteristic of the solution of Eqs.~(\ref{st LL2}). Using the
well-known connection between a set of stochastic differential equations and
the corresponding Fokker-Planck equation, see, e.g., Refs.~[\onlinecite{Gar,
Goel, HT, R}], we obtain the two-dimensional forward Fokker-Planck equation
\begin{eqnarray}
    \displaystyle\frac{\lambda}{2 a}
    \!\!\!&\bigg[&\!\!\!\!\frac{\partial^{2}P}{\partial\theta^{2}} +
    \frac{1} {\sin^{2}\theta}\frac{\partial^{2}P}{\partial\psi^{2}}
    \bigg] - \frac{\partial}{\partial\theta}\bigg[\frac{\lambda}{2 a}
    \cot\theta +  u(\theta,\psi) \bigg]P
    \nonumber\\[6pt]
    \displaystyle &-&\!\!\!\frac{\partial}{\partial\psi}
    [v(\theta,\psi) - \rho \Omega]P = \frac{\partial P}{\partial \tau}
    \label{fw FP}
\end{eqnarray}
and the corresponding backward Fokker-Planck equation
\begin{eqnarray}
    \displaystyle \frac{\lambda}{2a}
    \!\!\!&\bigg[&\!\!\!\!\frac{\partial^{2}P}{\partial\theta'^{2}} +
    \frac{1}{\sin^{2}\theta'}\frac{\partial^{2}P}{\partial\psi'^{2}}
    \bigg ] +
    \left[\frac{\lambda}{2 a} \cot\theta' + u(\theta',\psi')\right]
    \frac{\partial P}{\partial\theta'}
    \nonumber\\[6pt]
    \displaystyle &+&\!\!\![v(\theta',\psi') -
    \rho \Omega]\frac{\partial P}{\partial\psi'} =
    -\frac{\partial P}{\partial \tau'}.
    \label{bw FP}
\end{eqnarray}
Notably, this two-dimensional Fokker-Planck dynamics does not obey a detailed
balance symmetry,\cite{HT,R} if the driving frequency $\Omega$ does not vanish.
In the model under consideration we have $\tilde{W} = (1/2)\sin^{2}\theta -
\tilde{h} \sin\theta \cos\psi$, where $\tilde{h} = h/H_{a}$. Therefore
\begin{eqnarray}
    &u(\theta,\psi) = -\lambda\sin\theta\cos\theta + \tilde{h}\,
    (\lambda\cos\theta\cos\psi - \sin\psi),&
    \nonumber\\[6pt]
    &v(\theta,\psi) = \displaystyle \cos\theta - \tilde{h}\,
    \frac{\cos\theta\cos\psi + \lambda\sin\psi}{\sin\theta}.&
    \label{uv2}
\end{eqnarray}

The probability density $P$ must satisfy the equal-time condition $P|_{\tau =
\tau'} = \delta(\theta - \theta') \delta(\psi - \psi')$ and appropriate
boundary conditions, as implied by the physical context. Moreover, in spite of
the singularities in Eqs.~(\ref{fw FP}) and (\ref{bw FP}) at $\theta,\;\theta'
= 0,\pi$ (which are a consequence of the use of the spherical coordinate
system), the probability density $P$ must be a regular function also at these
points. In addition, if not excluded by the boundary conditions, $P$ must be
properly normalized, i.e., $\int_{0} ^{2\pi} d\psi \int_{0}^{\pi} d\theta \, P
= 1$. We note also that the forward and backward Fokker-Planck equations
(\ref{fw FP}) and (\ref{bw FP}) are equivalent; the difference between them is
which set of variables, $\theta,\psi, \tau$ or $\theta', \psi',\tau'$, is held
fixed. Due to this difference, the former is more convenient for studying the
statistical properties of the magnetic moment $\mathbf{m}(t)$ as functions of
the evolving time $t$, while the latter one is more appropriate in studying the
first-passage time statistics for $\mathbf{m}(t)$.

Based on the Fokker-Planck equation (\ref{fw FP}) one can determine
stochastically equivalent Langevin equations, reading
\begin{eqnarray}
    \dot{\theta} &=& u(\theta, \psi) + \frac{\lambda}{2a} \cot \theta +
    \sqrt{\frac{\lambda}{a}}\: \eta_{\theta}(\tau) \nonumber, \\
    \dot{\psi} &=& v(\theta, \psi) - \rho \Omega + \sqrt{\frac{\lambda}
    {a}}\:\frac{1}{\sin \theta} \:\eta_{\psi}(\tau), \label{LE2}
\end{eqnarray}
where $\eta_{\theta}(\tau)$ and $\eta_{\psi}(\tau)$ denote two independent
Gaussian white noise sources with zero mean and white noise correlations
$\langle \eta_i(\tau) \eta_j(\tau') \rangle = \delta_{ij} \delta(\tau -\tau')$,
wherein $i,j = \theta,\psi$. In spite of the multiplicative nature involving
the noise $\eta_{\psi}(\tau)$ in the second equation, the resulting stochastic
dynamics possesses a vanishing noise-induced drift and thus is again
independent of the employed stochastic calculus.\cite{Ber,BerG} As a basis for
numerical investigations the latter Langevin equations provide a more
convenient starting point than Eqs.~(\ref{st LL2}); this is so because they
require the simulation of only two, rather than three independent Gaussian
white noises.

\section{GENERAL EQUATIONS FOR THE MFPT$\textrm{s}$}

In the absence of the random magnetic field, i.e., for $\mathbf{n}(t) = 0$, or
equivalently $\Delta =0$, the motion of the magnetic moment follows the
deterministic Landau-Lifshitz equations. In the rotating frame the resulting
deterministic dynamical system is not explicitly dependent on time and given in
terms of  two degrees of freedom. For $\lambda>0$ this constitutes a
dissipative system, which can only perform regular motion approaching fixed
points or limit cycles in the asymptotic limit of large times. In the present
paper we are mainly interested in values of the parameters $\lambda$,
$\tilde{h}$ and $\Omega$  for which the motion is bistable, i.e., the
asymptotic motion does lead to either of two attractors depending on the
initial condition. These attractors are denoted as up and down states and
labelled  by $\sigma =+1$ and $\sigma = -1$, respectively. The dynamics
generates a partition of the state space into two domains of attraction,
containing either the up or the down state. The common boundary between the
domains of attraction is formed by the separatrix.

In the presence of a random magnetic field the separatrix is no longer an
impenetrable border, and transitions between the two domains of attraction may
occur. For small random fields, corresponding to large values of $a$, these
transitions are rare and can be characterized by transition rates. It is now
tempting to determine the rate between a state $\sigma$ and the opposite state
by the MFPT to the separatrix, $\mathcal{T}_{\sigma\: \text{sep}}$, i.e., by
the statistical average of the stochastic first-passage times of trajectories
that start at the attractor $\sigma$ and reach the separatrix  for the first
time. In the asymptotic limit of vanishing noise a trajectory visiting the
separatrix will go to either side with equal probability  and the rate is given
by the inverse of twice the mean first-passage time. For finite noise the
transition from the separatrix to the two sides may differ from each other,
\cite{ryter} whereby the precise value of the resulting bias in general is
difficult to quantify.\cite{talkner} In order to be more flexible, we consider
the mean first-passage times to the boundaries of two regions $R_{+1} =\{
\theta,\psi | 0 \leq \theta \leq \phi_{+1}(\psi), 0 \leq \psi < 2 \pi \}$, and
$R_{-1} =\{ \theta,\psi | \phi_{-1}(\psi) \leq \theta \leq \pi, 0 \leq \psi < 2
\pi \}$, each containing one attractor. The boundaries $\phi_{\sigma}(\psi)$
can be chosen as the separatrix or as any other curve between the two
attractors. For a convenient choice of the boundaries $\phi_\sigma(\psi)$ we
refer the reader to the next section. We note that the regions $R_\sigma$ are
stationary in the rotating frame but may move in the rest frame.

In order to determine the first-passage times of the regions $R_\sigma$,
re-crossings of the respective boundaries $\partial R_\sigma=\{\theta,\psi|
\theta=\phi_\sigma(\psi), 0\leq \psi \leq 2 \pi \}$ must be
suppressed.\cite{HTB} This is conveniently achieved by imposing absorbing
boundary conditions on the conditional probability density $P_\sigma(\theta,
\psi, \tau - \tau'|\theta', \psi')$ obeying the backward Fokker-Planck equation
(\ref{bw FP}), i.e., we require $P_\sigma(\theta,\psi,\tau-\tau'
|\theta',\psi') =0$ for $(\theta',\psi') \in \partial R_\sigma$. Here, we used
that in the rotating frame the process in the stationary regions $R_\sigma$ is
{\it time-homogeneous} with
\begin{eqnarray}
P_\sigma(\theta,\psi,\tau-\tau'|\theta',\psi') \!&\equiv&\!
P_\sigma(\theta,\psi,\tau-\tau'|\theta',\psi',0)
\nonumber \\
\!&=&\!P_\sigma(\theta,\psi,\tau|\theta',\psi',\tau') .
\end{eqnarray}

If the magnetic moment starts out at the time $\tau'$ at the position
$(\theta', \psi') \in R_\sigma$ it will uninterruptedly stays within the
initial region $R_\sigma$ until a time $\tau$ with a probability
$Q(\theta',\psi';\tau-\tau')$. This probability can be expressed as the
integral of the conditional probability density over all states in $R_\sigma$,
i.e.,
\begin{equation}
    Q_\sigma(\theta',\psi';\tau-\tau') = \int_{R_\sigma} d \psi d \theta\:
    P_\sigma(\theta,\psi, \tau-\tau'|\theta'\psi'). \;
\label{Q}
\end{equation}
The probability $Q_\sigma(\theta',\psi';\tau-\tau')$ is a solution of the
backward equation with the absorbing boundary conditions $Q_\sigma(\theta',
\psi';\tau-\tau') =0$ for $(\theta',\psi') \in \partial R_\sigma$ and the
initial condition $Q_\sigma(\theta',\psi';0) =1$. Integrating $Q_\sigma
(\theta', \psi';\tau-\tau')$ over all positive (dimensionless) times $u \equiv
\tau-\tau'$, one obtains an expression for the (dimensionless) MFPT of the form
\begin{equation}
    \mathcal{T}_{\sigma}(\theta',\psi') = \int_{0}^{\infty}du \,
    Q_{\sigma}( \theta',\psi';u).
    \label{mfpt1}
\end{equation}
This MFPT is the solution of the backward equation
\begin{eqnarray}
    \displaystyle\frac{\lambda}{2a} \!\!\!&\bigg[&\!\!\!\!
    \frac{\partial^{2}\mathcal{T}_{\sigma}}{\partial
    \theta'^{2}} + \frac{1}{\sin^{2}\theta'}\frac{\partial^{2}
    \mathcal{T}_{\sigma}}{\partial\psi'^{2}}\bigg ] +
    \bigg[\frac{\lambda}{2a}\cot\theta' +
    u(\theta',\psi')\bigg]\! \frac{\partial
    \mathcal{T}_{\sigma}}{\partial\theta'}
    \nonumber\\[6pt]
    \displaystyle &+&\!\!\! [v(\theta',\psi') - \rho \Omega]
    \frac{\partial \mathcal{T}_{\sigma}}{\partial\psi'} = -1
    \label{T sigma}
\end{eqnarray}
with the absorbing boundary conditions
\begin{equation}
\mathcal{T}_\sigma(\theta',\psi') = 0 \quad \text{for} \;(\theta',\psi')
\in \partial R_\sigma.
\label{bcMFPT}
\end{equation}
Eq.~(\ref{T sigma}) was derived in Ref.~[\onlinecite{DY}] for the undriven case
with $\Omega=0$.

Because $u (\theta', \psi')$, $v (\theta', \psi')$ and also $\mathcal{T}
_{\sigma} (\theta',\psi')$ are periodic functions of $\psi'$, it is
convenient to decompose these functions into their average and periodically
varying parts in $\psi'$: $u (\theta', \psi') = \overline{u}(\theta') + u_{1}
(\theta', \psi')$, $v (\theta', \psi') = \overline{v}(\theta') + v_{1}
(\theta', \psi')$, and $\mathcal{T}_{\sigma} (\theta', \psi') = \overline
{\mathcal{T}} _ {\sigma} (\theta') + \mathcal{S}_{\sigma} (\theta', \psi')$.
Here, $\overline{u}_{1} (\theta', \psi') = \overline{v}_{1} (\theta', \psi') =
\overline{\mathcal{S}} _{\sigma} (\theta', \psi') = 0$, the overbar denotes an
average over $\psi'$, i.e., $\overline {(\cdot)} = (1/2\pi) \int_{0}^{2\pi}
d\psi' (\cdot)$ and, according to (\ref{uv2}),
\begin{eqnarray}
    &\overline{u}(\theta') = -\lambda\sin\theta'\cos\theta', \quad
    \overline{v}(\theta') = \cos\theta',&
    \nonumber\\[6pt]
    &u_{1}(\theta',\psi') = \tilde{h}\,(\lambda\cos\theta'\cos\psi' -
    \sin\psi'),&
    \nonumber\\[6pt]
    &v_{1}(\theta',\psi') = \displaystyle - \tilde{h}\,
    \frac{\cos\theta'\cos\psi' + \lambda\sin\psi'}{\sin\theta'}.&
    \label{uu1 vv1}
\end{eqnarray}
Using these decompositions, we find from Eq.~(\ref{T sigma}) coupled equations,
see also in Ref.~[\onlinecite{DSTH}], for the average part
\begin{equation}
    \displaystyle\frac{\lambda}{2a} \bigg[\frac{d^{2}\overline
    {\mathcal{T}}_{\sigma}}{d\theta'^{2}} + \cot\theta'\frac{d
    \overline{\mathcal{T}}_{\sigma}}{d\theta'} \bigg ] \!
    + \overline{u} \frac{d \overline{\mathcal{T}}_
    {\sigma}}{d\theta'} + \overline{u_{1}\frac{\partial \mathcal{S}_{\sigma}}
    {\partial\theta'}} + \overline{v_{1}\frac{\partial \mathcal{S}_{\sigma}}
    {\partial\psi'}} = -1
    \label{over T1}
\end{equation}
and for the periodic part
\begin{eqnarray}
    \displaystyle\frac{\lambda}{2a} \!\!\!\!&\bigg[&\!\!\!\!
    \frac{\partial^{2}\mathcal{S}_{\sigma}}{\partial\theta'^{2}}
    \!+\! \frac{1}{\sin^{2}\theta'}\frac{\partial^{2}\mathcal{S}
    _{\sigma}}{\partial\psi'^{2}} + \cot\theta'\frac{\partial \mathcal{S}
    _{\sigma}}{\partial \theta'}\bigg ] \!\!+ u_{1}\frac{d
    \overline{\mathcal{T}}_{\sigma}} {d\theta'} + u\frac{\partial
    \mathcal{S}_{\sigma}}{\partial\theta'}
    \nonumber\\[6pt]
    \displaystyle &+&\!\!\! (v - \rho \Omega)\frac{\partial
    \mathcal{S}_{\sigma}}{\partial\psi'} - \overline{u_{1}\frac{\partial
    \mathcal{S}_{\sigma}} {\partial\theta'}} - \overline{v_{1}\frac{\partial
    \mathcal{S}_{\sigma}} {\partial\psi'}} = 0.
    \label{S1}
\end{eqnarray}
We emphasize that these equations are fully exact, i.e., they follow from the
stationary backward Fokker-Planck equation being in the rotating frame.

\section{RAPIDLY ROTATING FIELD}

\subsection{Analytical analysis for the MFPT}

In the case of a rapidly rotating magnetic field, i.e., when the condition
$\tilde{h} / \Omega \ll 1$ holds (note that $\tilde{h}$ need not be small),
Eq.~(\ref{S1}) can be essentially simplified.\cite{DSTH} The reason is that the
function $\mathcal{S} _{\sigma}$ and its derivatives tend to zero as $\tilde{h}
/ \Omega \to 0$. Using these conditions and taking into account that for large
frequencies the term $\Omega\, \partial \mathcal{S}_{\sigma}/ \partial \psi'$
is of the order $\Omega^{0}$,  we obtain from Eq.~(\ref{S1}) the approximation:
\begin{equation}
    \rho \Omega \frac{\partial \mathcal{S}_{\sigma}}{\partial \psi'} -
    u_{1}\frac{d\overline{\mathcal{T}}_{\sigma}}{d\theta'} = 0.
    \label{S3}
\end{equation}
According to (\ref{uu1 vv1}), the solution of Eq.~(\ref{S3}), which
satisfies the condition $\overline{\mathcal{S}}_{\sigma} = 0$, is
given by
\begin{equation}
    \mathcal{S}_{\sigma} = \rho \frac{\tilde{h}}{\Omega} (\lambda \cos \theta'
    \sin\psi' + \cos\psi') \frac{d\overline{\mathcal{T}}_{\sigma}}{d\theta'}.
    \label{sol S3}
\end{equation}
This solution self-consistently conforms with the assumptions made above. Using
(\ref{sol S3}) and (\ref{uu1 vv1}) we find
\begin{equation}
    \overline{u_{1}\frac{\partial \mathcal{S}_{\sigma}}{\partial \theta'}} =
    \overline{v_{1}\frac{\partial \mathcal{S}_{\sigma}}{\partial \psi'}} =
    - \frac{\lambda}{2}\tilde{h} _{\text{eff}} \sin \theta'
    \frac{d\overline{\mathcal{T}}_{\sigma}}{d\theta'},
    \label{over2}
\end{equation}
where
\begin{equation}
\tilde{h} _{\text{eff}} = -\rho \tilde{h}^{2} / \Omega .
\end{equation}
With these results, Eq.~(\ref{over T1}) reduces to the form
\begin{equation}
    \frac{\lambda}{2a}\frac{d^{2}\overline{\mathcal{T}}_{\sigma}}
    {d\theta'^{2}} + [\frac{\lambda}{2a}\cot\theta' - \lambda(\cos
    \theta' + \tilde{h} _{\text{eff}})\sin \theta']
    \frac{d \overline{\mathcal{T}}_{\sigma}}{d\theta'} = -1.
    \label{over T3}
\end{equation}

According to this equation, a magnetic field rapidly rotating in the plane
\textit{perpendicular} to the easy axis of the nanoparticle acts on the
nanoparticle's magnetic moment precisely as a \textit{static} effective
magnetic field $\tilde{h} _{\text{eff}}$ (in units of the anisotropy field
$H_{a}$) which is applied \textit{along} the easy axis. The direction of this
field and the direction of the field rotation follow the left-hand rule and its
value can be large enough to produce observable effects. Next we assume that
$|\tilde{h} _{\text{eff}}| < 1$; otherwise only one state, $\sigma = +1$ or
$\sigma = -1$, is stable.

It is not difficult to show that the general solution of Eq.~(\ref{over T3})
contains both regular and singular parts.\cite{DLT} The singular part arises
solely from the use of a spherical coordinate system and has no physical
meaning. It exhibits a logarithmic singularity at $\theta' = \pi(1-\sigma)/2$
and, as a consequence, its derivative diverges. On the contrary, the regular
part has a vanishing derivative at this point. Therefore, in order to exclude
the contribution of the singular part, the solution of Eq.~(\ref{over T3}) must
satisfy the regularity condition $(d\overline {\mathcal{T} }_{\sigma} /
d\theta') |_{\theta' = \pi(1-\sigma)/2} = 0$. We note in this context that the
regularity condition corresponds to the situation when a reflecting barrier is
placed at the point $\theta' = \pi(1-\sigma) /2$.

In order that the high frequency approximation for the MFPTs can consistently
be performed, the $\psi$ dependence of the boundary curves $\phi_{\sigma}
(\psi)$ must be chosen conveniently. Assuming that $\phi_{\sigma}(\psi') =
\overline {\phi}_{\sigma} + \phi_{1\sigma}(\psi')$ with $\overline
{\phi}_{1\sigma} (\psi') = 0$ and $\phi_{1\sigma}(\psi') \sim \tilde{h}/
\Omega$, the absorbing boundary condition (\ref{bcMFPT}) in the linear
approximation in $\tilde{h} / \Omega$ leads to the relations $\overline
{\mathcal{T}}_{\sigma} (\overline {\phi}_{\sigma}) = 0$ and $\phi_ {1{\sigma}}
(\psi') (d\overline {\mathcal{T}}_{\sigma} / d\theta')| _{\theta' = \overline
{\phi}_{\sigma}} + \mathcal{S}_{\sigma} (\overline {\phi}_{\sigma}, \psi') =
0$. Using (\ref{sol S3}) and the latter relation, we find for the absorbing
boundary the explicit form
\begin{equation}
    \phi_{\sigma}(\psi) = \overline {\phi}_{\sigma} - \rho \frac{\tilde{h}}
    {\Omega}(\lambda\cos\overline {\phi}_{\sigma}  \sin \psi + \cos \psi).
    \label{phi sigma}
\end{equation}

Next, solving Eq.~(\ref{over T3}) with the specified regularity and boundary
conditions, $(d\overline{\mathcal{T}}_{\sigma} / d\theta') |_{\theta' =
\pi(1-\sigma)/2} = 0$ and $\overline{\mathcal{T}}_{\sigma} (\overline
{\phi}_{\sigma}) = 0$, we obtain
\begin{equation}
    \overline{\mathcal{T}}_{\sigma}(\theta') = \frac{2a}{\lambda}\int_
    {\cos \overline {\phi} _{\sigma}}^{\cos\theta'}dx \,\frac{e^{-a(x +
    \tilde{h} _{\text{eff}})^2}}{1 - x^2} \int_{x}^{\sigma} dy\,
    e^{a(y + \tilde{h} _{\text{eff}})^2},
    \label{sol T3}
\end{equation}
where $\theta' \in [0, \overline {\phi}_{+1}]$ if $\sigma = +1$, and $\theta'
\in [\overline {\phi}_{-1}, \pi]$ if $\sigma = -1$. The angles $\overline
{\phi}_{\sigma}$ can be chosen depending on physical situation. For high
potential barriers, i.e.,  $a \gg 1$, the magnetic moment predominantly dwells
in the vicinity of either of two equilibrium states at $\theta = 0$ and $\theta
= \pi$. In this case the transition times between the states $\sigma$ and
$-\sigma$ exceed by far the relaxation times towards these states. Therefore,
in dimensional units the averaged MFPT $\overline{T} _{\sigma}(\theta') =
\overline{\mathcal{T}}_{\sigma} (\theta')/{\omega_r}$ representing the
transition time from one state $\sigma$ to the opposite state $-\sigma$ only
weakly depends on the precise location of the initial magnetization, as long as
$\theta'$ lies within the domain of attraction of the considered state
$\sigma$. Accordingly, the precise location of the absorbing boundary
$\overline {\phi}_ {\sigma}$ practically has no effect on $\overline{T}
_{\sigma} (\theta')$ if it is located well beyond the separatrix which divides
the state space into domains of attraction of the up and down magnetization.
Under these conditions,  we find  from (\ref{sol T3}) in leading order in $a$:
\begin{equation}
    \overline{T}_{\sigma} =  \frac{\overline{\mathcal{T}}_{\sigma}}
    {\omega_{r}} = \frac{1}{\lambda \omega_r} \sqrt{\frac{\pi}{a}}\,
    \frac{\exp{[a(1 + \sigma\tilde{h}_{\text{eff}})^2]}}{(1 - \tilde{h}
    _{\text{eff}}^2)(1 + \sigma \tilde{h} _{\text{eff}})}.
    \label{t6}
\end{equation}
As it follows upon inspection from (\ref{sol S3}) and (\ref{t6}), in the
high-frequency limit the periodic part of $T_{\sigma}(\theta',\psi')$ can be
neglected, i.e., $T_{\sigma}(\theta',\psi') \approx \overline{T}_{\sigma}$.

If $|\tilde{h}_{\text{eff}}| \ll 1$ then (\ref{t6}) yields $\overline{T}
_{\sigma} = T_{0} \exp(\sigma 2a \tilde{h}_{\text{eff}})$, where $T_{0} =
(1/\lambda \omega_r)\sqrt{\pi / a}\,\exp a$ is the MFPT at $\tilde{h} = 0$.
According to this formula, a rapidly rotating magnetic field increases the MFPT
for the magnetic moment in the state $\sigma = -\rho$ $(\sigma
\tilde{h}_{\text{eff}} > 0)$ and lowers this MFPT for the magnetic moment in
the state $\sigma = +\rho$ $(\sigma \tilde{h}_{\text{eff}} < 0)$. This
difference in the MFPTs arises from the natural precession of the nanoparticle
magnetic moments, which occurs in the counter-clockwise direction, if viewed
from above. As a consequence, the statistical behavior of the up and down
magnetic moments in the magnetic field rotating in a fixed direction is
different.

Another  choice of the position for the absorbing boundary can be made right on
the separatrix itself. For the averaged one dimensional dynamics of the
azimuthal angle it corresponds to the $\theta$ value where the deterministic
part of the drift of the reduced backward equation (\ref{over T3}) assumes an
unstable fixed point, i.e., for $\cos \theta = -\tilde{h}_{\text{eff}}$, or,
equivalently, to the maximum of the effective magnetic energy of the
nanoparticle, $W_{\text{eff}}(\theta) = mH_{a}[(1/2)\sin^2 \theta -
\tilde{h}_{\text{eff}} \cos \theta]$. Accordingly, the averaged dimensionless
MFPT from an initial angle $\theta'$ to the separatrix reads
\begin{equation}
    \overline{\mathcal{T}}_{\sigma\,\text{sep}}(\theta') =
    \frac{2a}{\lambda}\int_{-\tilde{h}
    _{\text{eff}}}^{\cos\theta'}dx \,\frac{e^{-a(x + \tilde{h}
    _{\text{eff}})^2}}{1 - x^2} \int_{x}^{\sigma} dy\, e^{a(y + \tilde{h}
    _{\text{eff}})^2}.
    \label{T sep}
\end{equation}
If $a \gg 1$ and $\cos \theta'$ is not too close to $-\tilde{h}_{\text{eff}}$
then $\overline{\mathcal{T}}_{\sigma\,\text{sep}}(\theta')$ only weakly depends
on $\theta'$ and $\overline{\mathcal{T}}_{\sigma\,\text{sep}} \to
\overline{\mathcal{T}}_{\sigma}/2$ as $a \to \infty$. Thus, the ratio
$\overline{\mathcal{T}}_{\sigma\,\text{sep}} / \overline{\mathcal{T}}_{\sigma}$
between the MFPT to the separatrix and to an angle $\overline {\phi}_{\sigma}$
which is well beyond the separatrix converges to the value $1/2$ if $a \to
\infty$; however, the larger the rotating field amplitude $\tilde{h}$ the
slower is the convergence, cf. in Fig.~1. We note also that for large but
finite $a$ the conditions $\overline{\mathcal{T}} _{\sigma\,\text{sep}} /
\overline{\mathcal{T}} _{\sigma} > 1/2$ and $\overline{\mathcal{T}}
_{\sigma\,\text{sep}} / \overline{\mathcal{T}} _{\sigma} < 1/2$ hold for
$\sigma \rho = +1$ and $\sigma \rho = -1$, respectively. The reason is that the
effective magnetic energy $W_{\text{eff}}(\theta)$ has different slopes from
the left and from the right of the separatrix disposed at $\theta = \arccos(-
\tilde{h}_{\text{eff}})$.

\subsection{Numerical simulations}

In order to examine the analytical results developed for calculating the MFPTs
for a magnetic moment which is driven by a rapidly varying circularly polarized
magnetic field, we performed numerical simulations of the full two-dimensional
Langevin equations (\ref{LE2}). For any given set of parameter values
$\rho,\sigma,\Omega,a,\lambda,\tilde{h}$ four groups of 10$^4$ trajectories
were run using a stochastic vector Euler algorithm. \cite{Gar, Gard} All
trajectories of a simulation were initialized with the same values for $\theta$
and $\psi$. In the vicinity of the coordinate singularity at $\theta = 0$ a
reflecting boundary was located at $\theta = 0.02 \pi$, together with an
absorbing boundary at $\theta = 0.8 \pi$, which is located well beyond the
separatrix of the corresponding deterministic system. The step width was chosen
such that the increments in $\theta$ were less than $\pi/100$ and the
increments in $\psi$  less than $\pi/10$, respectively. The precise computation
of the MFPTs of this system using the Langevin equations (\ref{LE2}) requires
to compute the arrival of all trajectories at the absorbing boundary. In
practice this method is unfeasible for all but the smallest values of the
anisotropy barrier height $a$; this is so because a given trajectory can take
much longer than the MFPT to arrive at the absorbing boundary. These events are
rare, but contribute significantly to the MFPT, and hence a sufficiently large
number of these events needs to be simulated to arrive at a reliable
statistics. However, assuming that the first-passage time distribution is
exponentially distributed with a rate parameter $1/\overline{T}_{\sigma}$ it is
possible to determine the MFPT approximately by fitting the tail of the
simulated first-passage time distribution to an exponential. Although this
assumption can be justified for large barrier heights $a \gg 1$ since then the
relaxation time of the system is much shorter than the MFPT, it clearly is
expected to fail for $a \approx 1$. Note, that $a$ enters the exponent of the
expression for the MFPT equation (\ref{sol T3}) and therefore, the large
barrier limit is already obtained for moderately large values of $a$. Using
this assumption allows to simulate the Langevin equations (\ref{LE2}) for each
trajectory up to a fixed maximal time at which on average a considerable
fraction of all trajectories, but not all, have crossed the absorbing boundary.
To be definite, we took this time to be two thirds of the theoretical mean
first-passage time calculated from equation (\ref{sol T3}). With this choice,
roughly half of all trajectories arrived at the absorbing boundary. The number
of absorbed transitions was stored as a function of time. From each of the
resulting four data sets the rate $1/\overline{T}_{\sigma}$ was determined by
an exponential fit and thereby the MFPT was estimated. From these four values
an average value and a standard deviation was determined. The so obtained
numerical findings compare most favorably with our theoretical predictions, cf.
Figs.~2 and 3.

Fig. 2 depicts the dependence of the MFPT as a function of the anisotropy
barrier height $a$ for two different driving angular frequencies $\Omega = 5$
and $\Omega = 10$. The values of the other parameters are $\rho = +1$, $\sigma
= +1$, $\tilde{h} = 1$, $\lambda = 0.1$. Depicted is the natural logarithm of
the dimensionless MFPT $\overline{\mathcal{T}}_{+1}$. It can clearly be seen
that the high-frequency predictions from the equations (\ref{sol T3}) and
(\ref{t6}) approach the results of the numerical simulation from above as
$\Omega$ is increased. This can be understood intuitively because the effective
dynamic barrier for escape is increased as $\Omega$ is increased. As expected,
the agreement decreases in quality upon lowering the angular driving frequency;
this fact is corroborated with the numerical results for $\Omega=5$, cf. the
thick dashed line versus the numerical data points.

Fig. 3 shows the dependence of the MFPT as a function of $\tilde{h}$, $\lambda$
and $\sigma$ at constant $a = 5$ and $\rho = +1$. As predicted by equations
(\ref{sol T3}) and (\ref{t6}), $\overline{\mathcal{T}}_{\sigma}$ can be seen to
be proportional to $1/\lambda$. The up and the down states ($\sigma = +1$ and
$\sigma = -1$) are clearly inequivalent for non-zero $\tilde{h}$. This is in
perfect agreement with the prediction of the theory, as can be deduced from the
approximate equation (\ref{t6}) in which $\sigma = +1$ ($\sigma = -1$)
decreases (increases) the exponent, since $\tilde{h}_{\text{eff}} < 0$ for
$\rho = +1$.

\section{RELAXATION OF THE MAGNETIZATION}

\subsection{Relaxation law at high anisotropy barrier}

The MFPTs $\overline{T}_{\sigma}$ provide important characteristics for the
magnetic dynamics of nanoparticles. If their states $\sigma$ are used for
information storage then the average times during which the information in
these states is kept safely must be considerably shorter than
$\overline{T}_{\sigma}$. The dependence of $\overline{T}_{\sigma}$ on the
characteristics of the rotating field gives a possibility to intentionally
change the relative stability of the up and down states.

The transition times $\overline{T}_{\sigma}$ from the state $\sigma$ to the
state $-\sigma$ also determine the thermally activated magnetic relaxation in a
system composed of uniaxial nanoparticles whose easy axes are perpendicular to
the plane of field rotation. If $a \gg 1$ and the precession angle\cite{DLHT}
\begin{equation}
    \theta_{\sigma} = \sqrt{\frac{1 + \lambda^2}{(1 -
    \sigma \rho \Omega)^2 + \lambda^2}}\, \tilde{h}
    \label{theta sigma}
\end{equation}
of the magnetic moment in the state $\sigma$ is small, i.e., $\theta_ {\sigma}
^{2} \ll 1$, the reduced magnetization of this system can be defined as $\mu(t)
= [N_{+1}(t) - N_{-1}(t)]/N$, where $N_{\sigma}(t)$ denotes the number of
nanoparticles in the state $\sigma$ at time $t$, and $N$ the whole number of
particles. Taking into account that $N_{+1}(t) + N_{-1}(t) = N$, this
definition yields $\dot{\mu}(t) = 2 \dot{N}_{+1}(t)/N$. Next, since the rate
$1/ \overline{T}_{\sigma}$ is the probability of reorientation of the magnetic
moment from the state $\sigma$ to the state $-\sigma$ per unit time, we have
$\dot{N}_{\sigma}(t) = N_{-\sigma}(t)/\overline{T}_{-\sigma} - N_{\sigma}(t)
/\overline{T}_{\sigma}$ and thus the equation for $\mu(t)$ takes the form
\begin{equation}
    \dot{\mu}(t) = -\mu(t)\left(\frac{1}{\overline{T}_{+1}} +
    \frac{1}{\overline{T}_{-1}}\right)
    - \frac{1}{\overline{T}_{+1}} + \frac{1}{\overline{T}_{-1}}.
    \label{eq mu}
\end{equation}
Its solution with the initial condition $\mu(0) = 1$ is given by
\begin{equation}
    \mu(t) = (1-\mu_{\infty})\exp(-t/t_{\text{rel}}) + \mu_{\infty},
    \label{sol mu}
\end{equation}
where $t_{\text{rel}} = \overline{T}_{+1}\overline{T}_{-1} / (\overline{T}_{+1}
+ \overline{T}_{-1})$ is the relaxation time and
\begin{equation}
    \mu_{\infty} = \frac{\overline{T}_{+1} - \overline{T}_{-1}}
    {\overline{T}_{+1} + \overline{T}_{-1}}
    \label{mu infty1}
\end{equation}
is the steady-state magnetization of the nanoparticle system induced by a
rapidly rotating field.

A simple analysis shows that a rapidly rotating magnetic field causes a
\textit{dec\-rease} of the relaxation time and \textit{magnetizes} the
nanoparticle system \textit{along} the easy axis of magnetization ($t_{\text
{rel}} < t_{\text{rel}}^{(0)}$ and $\mu_{\infty} \neq 0$ if $\tilde{h} \neq
0$). This conclusion is not trivial because the rotating field has no component
in the direction of the induced magnetization. The value of this magnetization
grows with decreasing temperature and its sign is determined by the direction
of the magnetic field rotation, i.e., $\textrm{sgn}\, \mu_{\infty} = -\rho$. In
particular, if $|\tilde{h} _{\text{eff}}| \ll 1$ then $t_{\text{rel}} =
(T_{0}/2) \cosh^{-1} (2a\tilde{h} _{\text{eff}})$ and $\mu_{\infty} = \tanh
(2a\tilde{h} _{\text{eff}})$. These results indicate that even a weak rotating
field can drastically decrease the relaxation time and strongly magnetize the
nanoparticle system if the temperature is low enough that $a|\tilde{h}
_{\text{eff}}| \gg 1$. We note also that Eq.~(\ref{mu infty1}) overestimates
the absolute value of the magnetization because the magnetic moment is less
than $m$ in the up state and larger than $-m$ in the down state.

\subsection{Steady-state magnetization at high frequencies}

In the case of a rapidly rotating field we are able to calculate the
steady-state magnetization $\mu_{\infty}$ for an arbitrary anisotropy barrier.
To this end, we introduce the stationary probability density $P_{ \text{st}} =
P_{\text{st}} (\theta,\psi)$ which, according to Eq.~(\ref{fw FP}), satisfies
the stationary (in the rotating frame) Fokker-Planck equation
\begin{eqnarray}
    \displaystyle\frac{\lambda}{2a} \!\!\!\!&\bigg[&\!\!\!\!
    \frac{\partial^{2}P_{\text{st}}}{\partial\theta^{2}}
    + \frac{1} {\sin^{2}\theta}\frac{\partial^{2}P_{\text{st}}}
    {\partial\psi^{2}}\bigg ] -
    \frac{\partial}{\partial\theta}\!\bigg[\frac{\lambda}{2a}\cot\theta
    +  u(\theta,\psi) \bigg]\!P_{\text{st}}
    \nonumber\\[6pt]
    \displaystyle &-&\!\!\!  \frac{\partial}{\partial
    \psi}[v(\theta,\psi) - \rho \Omega]P_{\text{st}} = 0
    \label{Pst}
\end{eqnarray}
being properly normalized, $\int_{0}^{2\pi}d\psi \int_{0}^{\pi} d\theta
P_{\text{st}} = 1$. Using the decompositions $P_{\text{st}} = \overline{P}_
{\text{st}} (\theta) + P_{1}(\theta,\psi)$ ($\overline{P}_{1} = 0$), $u
(\theta, \psi) = \overline{u}(\theta) + u_{1}(\theta, \psi)$ ($\overline{u}_{1}
= 0$), and $v (\theta, \psi) = \overline{v}(\theta) + v_{1} (\theta, \psi)$
($\overline{v}_{1} = 0$), where the overbar denotes averaging over $\psi$,
i.e., $\overline {(\cdot)} = (1/2\pi) \int_{0}^{2\pi}d\psi (\cdot)$,  we obtain
from Eq.~(\ref{Pst}) coupled equations for $\overline{P}_{\text{st}}$
\begin{equation}
    \frac{\lambda}{2a}\frac{d^2 \overline{P}_{\text{st}}}{d\theta^2} -
    \frac{d}{d\theta}
    \!\left[\frac{\lambda}{2a}\cot \theta + \overline{u} \right]\!
    \overline{P}_{\text{st}} - \frac{d}{d\theta}\;
    \overline{u_{1}P_{1}} =0
    \label{over Pst}
\end{equation}
and $P_{1}$
\begin{eqnarray}
    \displaystyle\frac{\lambda}{2a}\!\!\!\!&\bigg[&\!\!\!\!
    \frac{\partial^{2}P_{1}}{\partial\theta^{2}}
    + \frac{1} {\sin^{2}\theta}\frac{\partial^{2}P_{1}}
    {\partial\psi^{2}} \bigg ] -
    \frac{\partial}{\partial\theta}\!\bigg[\frac{\lambda}{2a} \cot\theta
    + u(\theta,\psi) \bigg]\!P_{1}
    \nonumber\\[6pt]
    \displaystyle &+&\!\!\! \rho \Omega
    \frac{\partial}{\partial\psi}P_{1} - \frac{\partial}{\partial\psi}
    \,v_{1}\overline{P}_{\text{st}} - \frac{\partial}{\partial\theta}
    \,u_{1}\overline{P}_{\text{st}} + \frac{d}{d\theta}\,
    \overline{u_{1}P_{1}}
    \nonumber\\[6pt]
    \displaystyle &-&\!\!\! \frac{\partial}{\partial\psi}\,
    v(\theta,\psi)P_{1} = 0.
    \label{P1a}
\end{eqnarray}

Assuming that $P_{1} \sim \Omega^{-1}$ as $\Omega \to \infty$,
Eq.~(\ref{P1a}) reduces in the high-frequency limit to
\begin{equation}
    \rho\Omega\frac{\partial}{\partial\psi}P_{1} = \frac{\partial}
    {\partial\psi}\;v_{1}\overline{P}_{\text{st}} + \frac{\partial}
    {\partial\theta}\; u_{1}\overline{P}_{\text{st}}.
    \label{P1}
\end{equation}
Since $v_{1} = - \tilde{h} (\cos\theta\cos\psi + \lambda\sin\psi)/\sin\theta$
and $u_{1} = \tilde{h} (\lambda\cos\theta\cos\psi - \sin\psi)$, the last
equation has the solution
\begin{eqnarray}
    \displaystyle P_{1}
    \!\!&=&\!\! -\rho \frac{\tilde{h}}{\Omega}(\cos\theta \cos\psi
    + \lambda\sin\psi) \frac{\overline{P}_{\text{st}}}{\sin\theta}
    \nonumber\\[6pt]
    \displaystyle &&\!\!+ \rho \frac{\tilde{h}}{\Omega}
    \frac{\partial}{\partial\theta}(\lambda\cos\theta \sin\psi
    + \cos\psi)\overline{P}_{\text{st}}.
    \label{sol P1}
\end{eqnarray}
Evaluating  the average
\begin{equation}
    \overline{u_{1}P_{1}} = \rho \frac{\lambda \tilde{h}^2}
    {\Omega}\sin\theta\; \overline{P}_{\text{st}},
    \label{aver}
\end{equation}
Eq.~(\ref{over Pst}) becomes
\begin{equation}
    \frac{\lambda}{2a}\frac{d^2
      \overline{P}_{\text{st}}}{d\theta^2} -
    \frac{d}{d \theta} \left [ \frac{\lambda}{2a}\cot \theta - \lambda
      (\cos\theta +
      \tilde{h} _{\text{eff}})\sin\theta \right ]
    \overline{P}_{\text{st}} =0. \;
    \label{over Pst2}
\end{equation}
The normalized solution of this equation assumes the form
\begin{equation}
    \overline{P}_{\text{st}}(\theta) = C \sin\theta \exp[-2a\tilde{W}_
    {\text{eff}}(\theta)],
    \label{sol Pst}
\end{equation}
where $\tilde{W} _{\text{eff}}(\theta) \equiv W _{\text{eff}}(\theta) / mH_{a}
= (1/2)\sin^2 \theta - \tilde{h} _{\text{eff}} \cos \theta$ is the
dimensionless effective energy of the nanoparticle and
\begin{equation}
    C = \sqrt{\frac{a}{\pi^3}} \frac{\exp[a(1 + \tilde{h}_{\text{eff}}^2)]}
    {\text{erfi}[\sqrt{a}(1 + \tilde{h}_{\text{eff}})] + \text{erfi}
    [\sqrt{a}(1 - \tilde{h}_{\text{eff}})]}
    \label{C}
\end{equation}
is a normalizing constant derived from the condition $2\pi \int_{0}^{\pi}
d\theta \overline{P}_{\text{st}} (\theta) = 1$. Here $\text{erfi}(z)$  stands
for the imaginary error function defined as $\text{erfi} (z) = (2/\sqrt {\pi})
\int_{0}^{z}dx \exp(x^2)$.

Finally, using the definition of the steady-state magnetization, $\mu_{\infty}
= 2\pi\int_{0}^{\pi} d\theta \cos\theta \overline{P}_ {\text{st}} (\theta)$, we
obtain
\begin{equation}
    \mu_{\infty} = \sqrt{\frac{1}{\pi a}} \frac{\exp[a(1 + \tilde{h}_
    {\text{eff}})^2] - \exp[a(1 - \tilde{h}_ {\text{eff}})^2]}
    {\text{erfi}[\sqrt{a}(1 + \tilde{h}_{\text{eff}})] + \text{erfi}
    [\sqrt{a}(1 - \tilde{h}_{\text{eff}})]} - \tilde{h}_{\text{eff}}.
    \label{mu infty2}
\end{equation}
Since $\text{erfi} (z) = 2(z + z^3/3 + \ldots)/ \sqrt{\pi}$ if $z
\ll 1$ and $\text{erfi} (z) = \exp(z^2)(1/z + 1/2z^3 + \ldots)/
\sqrt{\pi}$ if $z \gg 1$, in the case of low anisotropy barrier ($a
\ll 1$) this formula yields $\mu_{\infty} = 2a
\tilde{h}_{\text{eff}} / 3$, and in the case of high anisotropy
barrier ($a \gg 1$) and small effective field ($|\tilde{h}
_{\text{eff}}| \ll 1$) it is reduced to the formula $\mu_{\infty} =
\text{tanh}(2a \tilde{h} _{\text{eff}})$, which coincides with that
derived from the MFPT approach. As an illustration of the accuracy
of (\ref{mu infty2}) and the applicability of the MFPT approach for
describing the magnetic relaxation in nanoparticle systems, we
depict in Fig.~4 the theoretical and the numerical results for the
dependence of the induced magnetization $\mu_{\infty}$ on the
anisotropy barrier height $a= mH_a/2k_BT$.

\subsection{Numerical verification}

As in the simulation of the MFPTs, described in the previous section, the
Langevin equations (\ref{LE2}) were used to compute the mean magnetization at
high frequencies. However, the numerical simulation in this case differs from
the case studied before in two important aspects. Firstly, instead of an
ensemble average over magnetic nanoparticles a single, stochastic magnetic
moment was averaged over time, thereby making use of ergodicity. Secondly,
instead of using a reflecting and an absorbing boundary, here two reflecting
boundaries were located at $\theta = 0.01\pi$ and at $\theta = 0.99\pi$. For
fixed parameter values $\Omega = 10$, $\tilde{h} = 1$, $\rho = -1$, $\lambda =
0.5$ four trajectories were simulated for each value of $a$. Each trajectory
was initialized with $\theta = 0.05\pi$ and $\psi = \pi$. We introduced two
circles as marks on the sphere at $\theta = 0.2\pi$ and $\theta = 0.8 \pi$ and
defined a sign change of the magnetic moment as a crossing of the $\theta =
0.8\pi$ ($\theta = 0.2\pi$) mark, provided the magnetic moment had been in the
$\theta \leq 0.2 \pi$ ($\theta \geq 0.8\pi$) domain before. Each trajectory was
run until two hundred such sign changes had occurred. The projection of the
magnetic moment on the easy axis was summed over all time steps and divided by
the number of time steps at the end of the simulation. The convergence of the
mean magnetic moment along these trajectories was monitored and found to have
converged after two hundred sign changes. From the four values for the mean
magnetization a mean value and a standard deviation were determined. Fig.~4
depicts that the simulation and the analytic result of the steady-state theory
at high frequencies, equation (\ref{mu infty2}) are in very good agreement,
indicating that the high frequency limit is already obtained for $\Omega = 10$.

\section{CONCLUSIONS}

We carried out a comprehensive study of the two-dimensional MFPT problem for
the magnetic moment of a nanoparticle driven by a magnetic field rapidly
rotating in the plane perpendicular to the easy axis of magnetization. Our
approach is based on the equations (\ref{over T1}) and (\ref{S1}) for the MFPTs
that we derived from the backward Fokker-Planck equation in the rotating frame.
In the high-frequency limit, we solved these equations analytically and
calculated the MFPTs for the nanoparticle magnetic moment in the up and down
states. The main finding is that a rapidly rotating field influences the MFPTs
due to the change of the potential barrier between these states, which occurs
under the action of the static effective magnetic field applied along the easy
axis of magnetization. We showed that the magnetic field rotating in the
clockwise (counter-clockwise) direction increases the MFPT for the magnetic
moment in the up (down) state and decreases it for the magnetic moment in the
down (up) state. Our theoretical predictions are in good agreement with the
results obtained by numerical solution of the effective stochastic
Landau-Lifshitz equations.

In addition, we applied the derived MFPTs to study the features of magnetic
relaxation in nanoparticle systems caused by a rotating magnetic field. We
established that in the case of a large anisotropy barrier this field always
decreases the relaxation time and magnetizes the nanoparticle system. The
magnetization grows as the temperature decreases and its direction is uniquely
determined by the direction of field rotation. Solving the forward
Fokker-Planck equation in the case of a rapidly rotating field, we calculated
also the magnetization for an arbitrary anisotropy barrier. The theoretical
results are in excellent agreement with the numerical ones and confirm the
applicability of the MFPT approach for describing the magnetic relaxation in
systems of high-anisotropy nanoparticles driven by a rapidly rotating magnetic
field.

Due to the selective change of the noise-induced stability of a rapidly driven
magnetic moment of a nanoparticle, as represented by corresponding mean
first-passage times, the results herein can be used for potential applications
in magnetic recording technology. The relative stability of the magnetic moment
in the up and down states can be suitably controlled by either changing the
temperature $T$ of the environment (thereby changing the anisotropy barrier
height) or upon varying the strength of a rapidly rotating magnetic field.

\section*{ACKNOWLEDGMENTS}

S.I.D. acknowledges the support of the EU through contracts No
NMP4-CT-2004-013545 and No MIF1-CT-2006-021533, P.T. and P.H. acknowledge the
support of the DFG via the SFB 486, project A 6.

\clearpage

\begin{figure}
    \centering
    \includegraphics[totalheight=5cm]{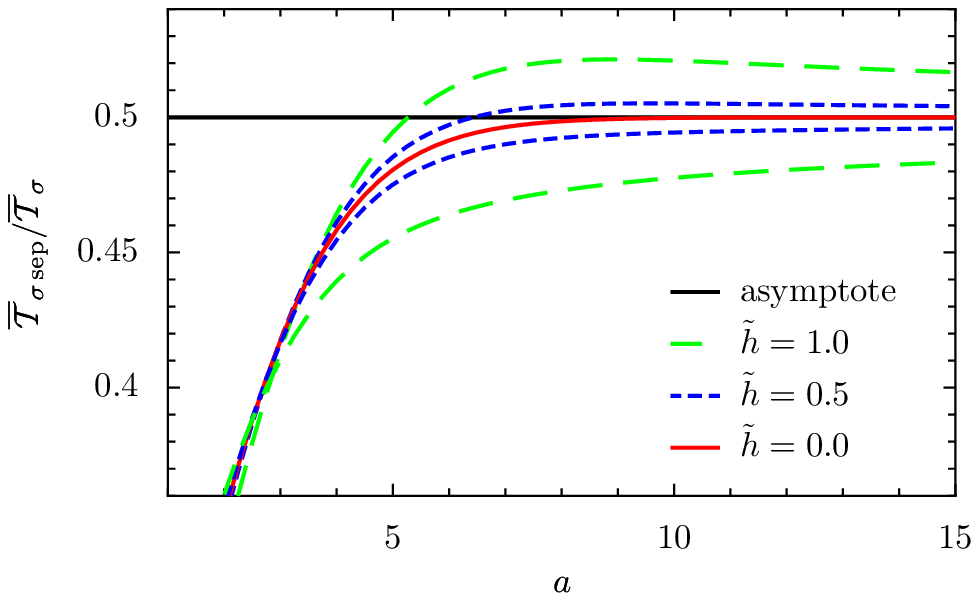}
    \caption{Dependence of the ratio $\overline{\mathcal{T}}_{\sigma\,
    \text{sep}} / \overline{\mathcal{T}}_{\sigma}$
    on the dimensionless anisotropy barrier height $a$
    for different values of the dimensionless amplitude $\tilde{h}$
    of the rotating magnetic field. The numerical calculations of the
    integrals in the relations (\ref{sol T3}) and (\ref{T sep}) were
    carried out for $\Omega = 10$, $\rho = +1$, $\overline {\phi}_
    {+1} = 0.9\pi$, $\overline {\phi} _{-1} = 0.1\pi$, $\theta' =
    0.1\pi$ if $\sigma = +1$, and $\theta' = 0.9\pi$ if $\sigma =
    -1$. The broken curves (green and blue online) that cross the
    horizontal asymptote (black online) correspond to $\sigma = +1$,
    and the broken curves (green and blue online) that lie below the
    asymptote correspond to $\sigma = -1$. The solid curve (red
    online) represents the case $\tilde{h} = 0$ for which $\overline
    {\mathcal{T}}_{+1\,\text{sep}} / \overline{\mathcal{T}}_{+1} =
    \overline {\mathcal{T}}_{-1\,\text{sep}} / \overline{\mathcal{T}}
    _{-1}$. We note also that in all cases the ratio $\overline
    {\mathcal{T}}_{\sigma\,\text{sep}} / \overline{\mathcal{T}}_
    {\sigma}$ does not depend on $\lambda$.}
\end{figure}

\clearpage

\begin{figure}
    \centering
    \includegraphics[totalheight=5cm]{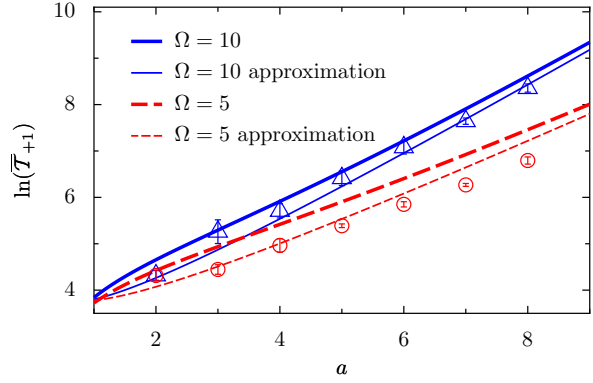}
    \caption{The natural logarithm of the dimensionless MFPT
    $\overline{\mathcal{T}}_{+1}$ as a function of the
    parameter $a$. The thick curves represent the exact theoretical
    results obtained from (\ref{sol T3}), and the thin curves depict
    the approximate high barrier limit given by (\ref{t6}). The symbols
    indicate results from the numerical simulation of $4\times 10^{4}$
    runs of the effective stochastic Landau-Lifshitz equations
    (\ref{LE2}) with the initial conditions $\psi_{0} = 0$,
    $\theta_{0} = 0.05\pi$ and with the absorbing
    boundary at $\overline{\phi}_{+1} = 0.8\pi$.
    The broken curves and circular symbols (red online) correspond to
    $\Omega = 5$, and the solid curves and triangular symbols (blue
    online) correspond to $\Omega = 10$. In accordance with the
    theoretical assumption, the analytical results approach to the
    numerical ones with increasing of the field frequency.}
\end{figure}

\clearpage

\begin{figure}
    \centering
    \includegraphics[totalheight=5cm]{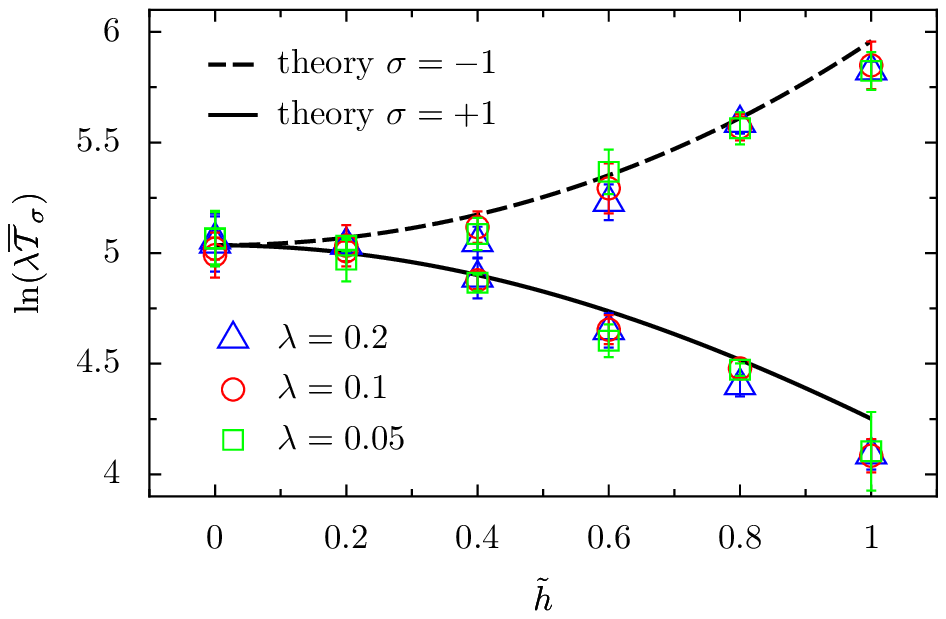}
    \caption{The natural logarithm of $\lambda \overline{\mathcal{T}}
    _{\sigma}$ as a function of the dimensionless amplitude $\tilde{h}$
    of the rotating magnetic field. The solid and broken curves represent
    the theoretical results obtained from the relation
    (\ref{sol T3}) for $\sigma = +1$ and $\sigma = -1$,
    respectively. The values of the other parameters are $a = 5$,
    $\Omega = 10$, and $\rho = +1$. The symbols (in color online)
    depict the results obtained from the numerical simulation of
    Eqs.~(\ref{LE2}) for different values of the damping
    parameter $\lambda$. In full agreement with theoretical
    predictions, $\ln (\lambda \overline{\mathcal{T}}_{\sigma})$ as a
    function of $\tilde{h}$ decreases if $\sigma = +1$, increases if
    $\sigma = -1$, and does not depend on $\lambda$.}
\end{figure}

\clearpage

\begin{figure}
    \centering
    \includegraphics[totalheight=5cm]{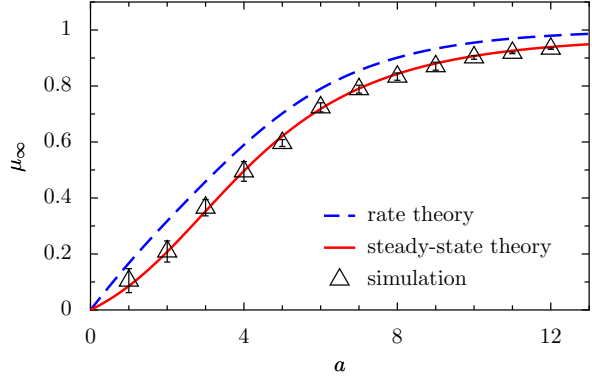}
    \caption{The dimensionless induced magnetization $\mu_{\infty}$
    as a function of the anisotropy barrier height $a$ for $\Omega = 10$,
    $\tilde{h} = 1$, $\rho = -1$, and $\lambda = 0.5$. The solid curve (red
    online) and the broken curve (blue online) represent the induced
    magnetization defined by the formulas (\ref{mu infty2}) and
    (\ref{mu infty1}) with (\ref{sol T3}), respectively. The
    triangular symbols indicate results obtained from the numerical
    simulation of Eqs.~(\ref{LE2}). As seen, the theoretical results
    that follow from the explicit solution of the Fokker-Planck equation
    are in excellent agreement with the numerical results. A small systematic
    shift of the induced magnetization derived within the rate theory arises
    from an overestimation of the absolute values of the average magnetic
    moment in the up and down states.}
\end{figure}

\end{document}